\documentclass[journal,11pt]{IEEEtran}
\IEEEoverridecommandlockouts

\usepackage[utf8]{inputenc}
\usepackage[T1]{fontenc}
\usepackage{microtype}
\usepackage{lmodern}
\usepackage{amsmath,amssymb}
\usepackage{graphicx}
\usepackage{booktabs}
\usepackage{tabularx}
\usepackage{array}
\usepackage{caption}
\usepackage{textgreek}
\usepackage{enumitem}
\usepackage{url}
\usepackage{hyperref}
\usepackage{textcomp}
\usepackage{footmisc}

\setlist[itemize]{noitemsep, topsep=4pt, left=1.5em}

\title{Instant Preliminary Cardiac Analysis from Smartphone Auscultation: A Real-World Canine Heart Sound Dataset and Evaluation
}

\author{
\IEEEauthorblockN{
\textbf{Aswin Jose}\IEEEauthorrefmark{1},
\textbf{Dr.\ Roeland P-J E.\ Decorte}\IEEEauthorrefmark{1},
\textbf{Laurent Locquet}\IEEEauthorrefmark{1}\IEEEauthorrefmark{2}
} \\
\IEEEauthorblockA{\IEEEauthorrefmark{1}
Decorte Future Industries Ltd / Sonus Health}
\\
\IEEEauthorblockA{\IEEEauthorrefmark{2}
Honorary Associate Professor, University of Nottingham}
}

\begin{document}
\maketitle

\begin{abstract}
This study presents a real-world canine heart sound dataset and evaluates SoNUS 3.2.x, a machine learning algorithm for preliminary cardiac analysis using smartphone microphone recordings. Over a hundred recordings were collected from dogs across four continents, with 38 recordings annotated by board-certified veterinary cardiologists for quantitative evaluation.

SoNUS 3.2.x employs a multi-stage fallback architecture with quality-aware filtering to ensure reliable output under variable recording conditions. The primary 60-second model achieved mean and median heart rate accuracies of 91.63\% and 94.95\%, while a fast model optimized for 30–40 second recordings achieved mean and median accuracies of 88.86\% and 92.98\%. These results demonstrate the feasibility of extracting clinically relevant cardiac information from opportunistic smartphone recordings, supporting scalable preliminary assessment and telehealth applications in veterinary cardiology.
\end{abstract}


% ---------- Sections (original text preserved) ----------
\section{Introduction}

Cardiovascular disease is one of the most common health concerns in companion animals, both in dogs and cats. Early detection and continuous monitoring are critical for improving clinical outcomes and overall quality of life. Furthermore, a wide range of systemic diseases and physiological states, including conditions associated with pain or stress, can exert measurable effects on cardiac parameters. Consequently, these relationships indicate that cardiac analysis may serve as a valuable, non-invasive adjunct for the triage and preliminary assessment of systemic health states, with the potential to identify physiological changes associated with non-cardiac disease processes.

In veterinary practice, cardiac assessment typically relies on auscultation of the thorax performed with a stethoscope, and when cardiovascular disease is suspected, tests such as echocardiography and electrocardiography are recommended. While effective, these methods are limited to clinical environments, require operator expertise, are time and resource intensive, and are not easily scalable for routine or at-home monitoring.

The prevalence of cardiovascular disease is significant in dogs, with myxomatous mitral valve disease (MMVD) accounting for up to 75\% of all cardiac disease cases in primary care. This condition affects both quality of life and life expectancy. Recent literature supports a more stepwise, Spectrum of Care (SoC) approach for the diagnosis and staging of MMVD, emphasising auscultation as a cornerstone of clinical assessment \cite{b9b6e34620144f408984adf57a6bf663}.

Temporal clinical and radiographic changes described by Boswood et al. further highlight the prognostic relevance of heart rate progression in preclinical MMVD, while murmur intensity plays a key role in identifying dogs suitable for medical treatment under the EPIC trial criteria \cite{Boswood2016EPIC,Boswood2020Temporal}.

In cats, cardiovascular disease presents distinct diagnostic and management challenges, with hypertrophic cardiomyopathy (HCM) representing the most commonly reported cardiac condition. In cases of hypertrophic obstructive cardiomyopathy (HOCM), pharmacological management in affected cats can include β-adrenergic blockade, such as atenolol, aimed at moderating heart rate, reducing myocardial oxygen demand, and alleviating dynamic outflow tract obstruction where present. 

The clinical response to atenolol is highly dependent on appropriate dosing and timing, as both under- and over-suppression of heart rate may adversely affect haemodynamic stability and overall clinical status. Consequently, longitudinal monitoring of cardiac parameters, particularly heart rate trends and rhythm stability, is important for evaluating therapeutic response and supporting ongoing dose optimisation. However, practical constraints on frequent in-clinic assessment limit the feasibility of such monitoring in routine feline care, underscoring the broader value of scalable, non-invasive approaches to repeated cardiac evaluation outside the clinical setting.

Sonus Health's artificial intelligence enables cardiac scanning using hardware already present in most households, specifically a smartphone’s microphone, supporting a scalable approach to routine or at-home cardiac monitoring, without reliance on specialised clinical environments or user expertise. The scanning process is structured as a two-stage workflow. In the first stage, automated preliminary results are generated immediately following acquisition of a recording. In the second stage, a comprehensive Full Analysis Report (FAR) is delivered within 48 hours using a hybrid framework that combines an advanced artificial intelligence analysis pipeline with human input, oversight and sign-off, provided by a board-certified veterinary cardiologist.

High-quality, real-world datasets for digital heart sound analysis are virtually non-existent, limiting the development and validation of algorithms capable of reliably detecting heart sounds and identifying cardiac abnormalities. The ability to create such algorithms is a prerequisite for the fully automated, real-time preliminary analysis stage of the Sonus Health approach.

To enable the development and validation of a real-time preliminary analysis system under real-world conditions, a large dataset of canine cardiac recordings was collected from diverse geographic regions. This study describes the dataset, recording, collection and annotation protocols, and presents quantitative performance results generated using the SoNUS (Sound-based Natural Understanding System) preliminary analysis pipeline.

This study makes three primary contributions. First, it introduces a real-world dataset of canine cardiac auscultation recordings acquired using smartphone microphones under routine clinical and at-home conditions, addressing a major gap in available veterinary heart sound data. Second, it presents a robust, multi-stage fallback architecture with integrated quality gating that enables reliable preliminary cardiac analysis despite substantial variability in recording quality. Third, it quantitatively evaluates heart rate estimation performance across both standard and short-duration recordings, demonstrating the feasibility of scalable, automatic and non-invasive cardiac triage and longitudinal monitoring in the veterinary space.

\section{Data}

Heart sound recordings were collected in real-world conditions from over a hundred dogs at various veterinary centers across 5 continents. The recordings were taken while the animals were in a calm resting state, with informed owner consent, and were collected in accordance with routine veterinary care practices without alteration to clinical decision-making or animal welfare. The dataset includes both healthy dogs and dogs with clinically diagnosed cardiac conditions, including murmurs, ectopic beats and arrhythmias, spanning a broad range of breeds, body sizes, and ages representative of general veterinary practice.

Recordings were performed by different veterinarians and nurses using their mobile phones, with the microphone placed against the thorax, reflecting realistic variability in recording quality and placement.

From the total pool, 38 recordings were manually annotated by a board-certified veterinary cardiologist and their team. Annotations include the timing of S1 and S2 heart sounds, the presence of murmurs, ectopic beats, and arrhythmias. Quantitative evaluation in this study focuses primarily on heart rate estimation due to its robustness to annotation variability and consistency across the dataset. Timestamp accuracy carries an estimated \(\pm 10\) ms error margin, consistent with human annotation limitations.

Each data point consists of an audio file and a corresponding annotation text file.

To evaluate model performance on shorter-duration inputs, we additionally generated a subset of short recordings by subsampling the original files. Three segments of 30–40 seconds with random overlaps were extracted from each annotated recording, resulting in a total of 114 short-duration recordings. This approach preserves physiological variability while enabling systematic evaluation of model variants optimized for shorter input lengths.

\section{Model}

The original version of the algorithm, \textbf{SoNUS 3.1.x}, was designed as a single-pass heart sound analysis pipeline primarily focused on heart rate detection. Given an input recording, the model processed the signal once, using adaptive heuristics based on factors such as background noise levels, recording quality, and expected physiological constraints of the organism. This version performed reliably when provided with high-quality recordings, but was sensitive to variations in input conditions.

During real-world deployment, particularly with mobile phone recordings, we observed substantial variability in audio characteristics caused by differences in device hardware, firmware, microphone placement, recording environments and user handling. These variations often resulted in otherwise valid recordings failing to meet the quality thresholds of SoNUS 3.1.x.

To address this limitation, we developed \textbf{SoNUS 3.2.x}, which extends the original SoNUS 3.1.x pipeline with a multi-stage fallback mechanism. In this version, the same core algorithm is applied iteratively across up to three stages. Each stage progressively relaxes quality thresholds and tolerance parameters, allowing the system to reprocess lower-quality recordings and recover valid results when possible. In the worst case, a recording may be processed three times instead of once, with each stage attempting to detect periodic heart sound patterns under increasingly permissive conditions.

Successful detection at Stage 1 typically indicates high-quality input and higher model confidence, whereas detections achieved at later stages reflect increased robustness to recording variability, at the cost of more restricted confidence in interpretation. We additionally introduced a quality scoring module that assigns a composite reliability score based on processing stage depth, internal confidence metrics, and the consistency of valid detections across audio windows.

A dedicated model variant optimized for shorter-duration recordings was also developed. While longer recordings provide additional context that improves result consistency and validation, the primary SoNUS pipeline was optimized for 60-second inputs based on available computational resources, processing time constraints and window-level consistency checks. Applying this configuration to shorter recordings (e.g., 30 seconds) can lead to suboptimal window alignment and reduced reliability of consistency metrics. To address this, we introduced a simplified model variant that retains the same core architecture but employs modified configurations tailored for shorter inputs in the 30–40 second range. This variant supports the rapid analysis functionality of SoNUS, while maintaining consistency with the primary model’s detection logic.

Preliminary analysis is performed using a computationally constrained pipeline designed for rapid processing of short-duration recordings, consistent with deployment in mobile or near-real-time application settings.

\begin{figure}[htbp]
    \centering
    \includegraphics[width=\linewidth]{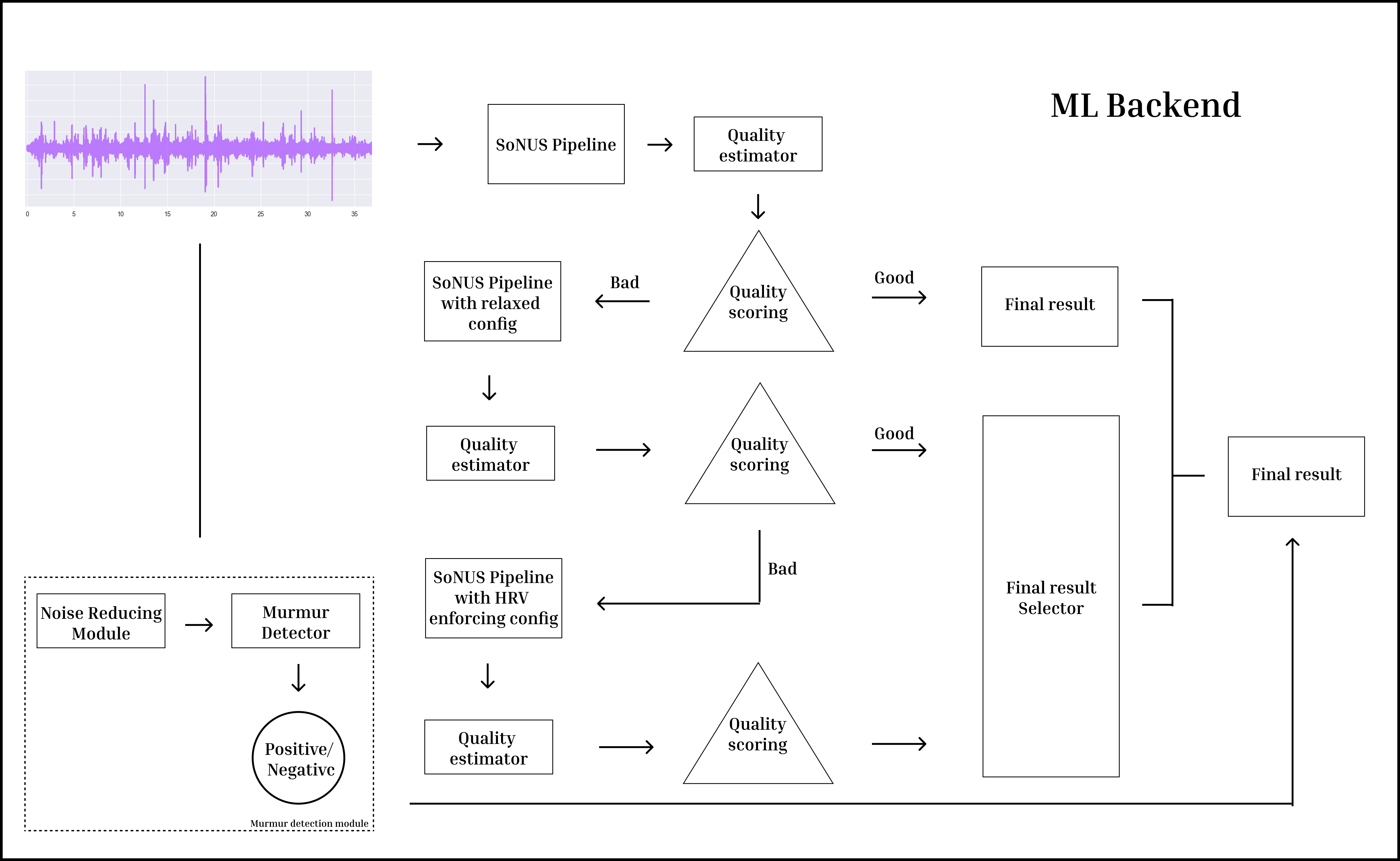}
    \caption{High-level overview of the SoNUS 3.2.x processing pipeline}
    \label{fig:myimage}
\end{figure}

In addition to heart rate estimation, we developed a murmur detection model based on a ResNet architecture. The model was initially trained and evaluated on the PhysioNet/CinC 2016 and 2022 challenge datasets, achieving an F1-score of 0.89 on the respective test sets, outperforming the reported state-of-the-art methods from the corresponding challenges. Building upon this already state-of-the-art foundation, the model was subsequently fine-tuned for canine cardiac auscultation using domain-specific data to adapt to differences in physiology, heart sound characteristics, and recording conditions.
\section{Results}

Model performance was evaluated using heart rate (HR) estimation error, reported as percentage error relative to cardiologist-annotated ground truth. Accuracy is expressed as $(100 - \text{HR error \%})$, a formulation chosen to provide an intuitive percentage-based measure of agreement with cardiologist-annotated ground truth. Results are presented separately for the fast (short-duration) model and the primary SoNUS model.

\subsection{Fast Model Performance (Short Recordings)}

The fast model, optimized for 30–40 second recordings, demonstrated robust performance despite reduced temporal context. Across all evaluated short recordings, the mean HR error was 11.14\%, corresponding to a mean HR accuracy of 88.86\%. The median HR error was 7.02\%, yielding a median HR accuracy of 92.98\%. These results indicate that the simplified configuration maintains reliable HR estimation for rapid analysis scenarios.

\begin{table}[h]
\centering
\caption{Fast Model HR Accuracy Percentiles (All Recordings)}
\begin{tabular}{lc}
\toprule
\textbf{Percentile} & \textbf{HR Accuracy (\%)} \\
\midrule
80th & 83.92 \\
90th & 74.04 \\
\bottomrule
\end{tabular}
\label{tab:fast_model_all}
\end{table}

To assess the effectiveness of the quality scoring module, results were further analyzed by filtering recordings with a quality score (QS) $\geq 70$. This threshold represents the minimum confidence level at which the application allows the presentation of results to end users. Recordings below this QS threshold are intentionally excluded due to insufficient signal reliability.

When filtering the short-recording dataset ($n=114$) by quality score, 98 recordings (86\%) had a QS $\geq 70$ and were deemed suitable for result presentation within the application. This indicates that the quality scoring module effectively identifies recordings with sufficient signal integrity, while excluding low-confidence inputs: a key requirement for real-world conditions and deciding when to show results to end users.

\begin{table}[h]
\centering
\caption{Fast Model HR Accuracy Percentiles (QS $\geq$ 70)}
\begin{tabular}{lc}
\toprule
\textbf{Percentile} & \textbf{HR Accuracy (\%)} \\
\midrule
80th & 84.82 \\
90th & 76.49 \\
\bottomrule
\end{tabular}
\label{tab:fast_model_qs70}
\end{table}

\subsection{Primary Model Performance (60-Second Recordings)}

The primary SoNUS model, optimized for 60-second recordings, achieved improved performance due to increased temporal context and stronger consistency checks. Across all annotated recordings ($n=38$), the mean HR error was 8.37\%, corresponding to a mean HR accuracy of 91.63\%. The median HR error was 5.05\%, yielding a median HR accuracy of 94.95\%.

When filtering by quality score, 32 out of 38 recordings (84.21\%) had a QS $\geq 70$ and were deemed suitable for result presentation within the application.

\begin{table}[h]
\centering
\caption{Primary Model HR Accuracy Percentiles (All Recordings)}
\begin{tabular}{lc}
\toprule
\textbf{Percentile} & \textbf{HR Accuracy (\%)} \\
\midrule
80th & 87.15 \\
90th & 78.88 \\
\bottomrule
\end{tabular}
\label{tab:main_model_all}
\end{table}

\begin{table}[h]
\centering
\caption{Primary Model HR Accuracy Percentiles (QS $\geq$ 70)}
\begin{tabular}{lc}
\toprule
\textbf{Percentile} & \textbf{HR Accuracy (\%)} \\
\midrule
80th & 88.71 \\
90th & 86.29 \\
\bottomrule
\end{tabular}
\label{tab:main_model_qs70}
\end{table}

\subsection{Impact of Quality Scoring}

Filtering results using the quality score threshold consistently improved percentile-level accuracy while reducing exposure to unreliable predictions. This demonstrates that the quality scoring module provides a meaningful proxy for signal reliability and model confidence. By restricting user-facing outputs to recordings with QS $\geq 70$, the system prioritizes result consistency and clinical interpretability over coverage, aligning with safety-oriented deployment requirements.
\subsection{Murmur Detection}

A comprehensive quantitative evaluation of the murmur detection model on the collected real-world dataset was not feasible, as murmur classification is highly sensitive to signal quality, microphone placement, and ambient noise (signal-to-noise ratio) - factors that vary substantially in opportunistic mobile phone recordings. As a result, model behavior in this setting is strongly influenced by the quality of the input data, and performance metrics must be interpreted in the context of recording reliability rather than algorithmic capability alone.

These observations highlight the need for larger, higher-fidelity real-world datasets for robust murmur classification and motivate ongoing data collection and targeted evaluation under controlled acquisition protocols.

\section{Discussion}

The results demonstrate that SoNUS 3.2.x achieves reliable heart rate (HR) estimation across real-world canine recordings collected under highly variable conditions. For the primary 60-second model, mean and median HR accuracies of 91.63\% and 94.95\%, respectively indicate strong central performance when sufficient temporal context is available. The fast model, optimized for shorter 30–40 second recordings, shows slightly reduced but still robust performance, with mean and median HR accuracies of 88.86\% and 92.98\%. This expected trade-off reflects the reduced signal context available for window-level consistency checks, while still supporting rapid analysis use cases. 

Heart rate estimates produced by the SoNUS pipeline were evaluated directly against manual heart rate annotations provided by board-certified veterinary cardiologists, representing a clinically grounded auscultation-based reference standard. While the number of cardiologist-annotated recordings is necessarily limited by the availability of expert annotation, the dataset reflects real-world acquisition conditions and prioritises ecological validity over controlled laboratory scale.

Importantly, the quality scoring module plays a critical role in ensuring reliability. Across both model variants, filtering results using a quality score (QS) threshold of 70 consistently improves percentile-level accuracy and suppresses low-confidence predictions. In practice, this results in 86\% of short recordings and 84.21\% of full-length recordings being deemed suitable for presentation within the application. These findings indicate that the quality score serves as an effective proxy for signal integrity and model confidence, allowing the system to balance coverage with interpretability and safety.

The multi-stage fallback design further enhances robustness to real-world variability. While the original SoNUS 3.1.x pipeline was limited to high-quality recordings, SoNUS 3.2.x successfully recovers valid results from lower-quality inputs without compromising performance on clean data. Successful detection at earlier stages correlates with higher confidence, whereas later-stage detections provide controlled tolerance to noise and recording artefacts. This design is particularly relevant for recordings obtained by external veterinary personnel or pet owners, where consistency of acquisition cannot be guaranteed.

Heart rate variability (HRV) estimation remains inherently more challenging than HR estimation due to noise sensitivity and the instability of short-window measurements. Rather than reporting absolute HRV values, SoNUS 3.2.x provides smoothened trend-based HRV representations, which are less sensitive to transient artefacts and more clinically interpretable in longitudinal monitoring scenarios.

While the murmur detection model demonstrated state-of-the-art and best-in-field benchmark performance on established PhysioNet challenge datasets\cite{PhysioNet2016, PhysioNet2022}, comprehensive quantitative evaluation on the collected real-world dataset was limited by recording quality variability and signal-to-noise constraints.

\section{Conclusion}

This initial study demonstrates the feasibility of using smartphone-based audio recordings combined with SoNUS AI for canine cardiac analysis. The data of over a hundred dog heart recordings, and specifically the 38 recordings further annotated by veterinary cardiologists, provide a reliable foundation for validating algorithm performance under real-world conditions.

SoNUS 3.2.x successfully estimates heart rate (HR) with high accuracy across variable recording environments. For the primary model optimized for 60-second recordings, mean and median HR accuracies of 91.63\% and 94.95\% demonstrate strong central performance when sufficient temporal context is available. The fast model, designed for shorter 30–40 second recordings, achieves mean and median accuracies of 88.86\% and 92.98\%, supporting rapid analysis use cases while maintaining reliable estimation performance. In addition to accurate heart rate estimation, the system supports trend-based heart rate variability monitoring suitable for longitudinal use under real-world recording conditions.

The multi-stage fallback design improves resilience to heterogeneous recording quality by enabling recovery of valid results from lower-quality inputs without compromising performance on high-quality data. In combination with the quality scoring module, which restricts user-facing outputs to recordings with a quality score of 70 or higher, the system prioritizes result reliability and interpretability. First-stage results collected by external veterinary personnel further demonstrate applicability beyond controlled or developer-driven recording settings.

These findings confirm that high-value cardiac information can be extracted from mobile microphone audio, supporting future development of telehealth and at-home monitoring solutions in veterinary cardiology.

From a veterinary clinical perspective, tools capable of accurately analysing heart sounds at scale can significantly complement traditional diagnostic pathways. Such tools may support earlier detection of cardiac disease, support Spectrum of Care models in general practice, facilitate longitudinal monitoring of medical therapy and medication management — including assessment of therapeutic response and the potential need for dose adjustment — and allow low-cost monitoring options that bridge the gap between auscultation and advanced imaging modalities.

For more complex rhythm disorders, such as atrial fibrillation in dogs, this type of scalable acoustic monitoring can offer a more accessible and cost-effective alternative to prolonged ambulatory electrocardiographic monitoring in settings where 24-hour Holter monitoring is unavailable or impractical. In dogs with myxomatous mitral valve disease (MMVD) and congestive heart failure, longitudinal trends in heart rate variability can provide additional insight into physiological stability and treatment response when assessed in the home environment, with higher HRV values commonly observed in dogs with well-controlled disease. Additionally, at-home identification of murmur-related acoustic features may help flag cases that warrant earlier or more detailed clinical investigation than might otherwise occur through intermittent in-clinic assessment alone.

While the present study focuses on canine recordings, the SoNUS system is currently used in routine veterinary clinic contexts for both dogs and cats. Ongoing work includes expansion of annotated canine and feline datasets and the collection of a large equine heart sound dataset, supporting future evaluation of scalability across species and veterinary use cases.

\bibliographystyle{IEEEtran}
\bibliography{references}
% ---------- End ----------
\end{document}